\documentclass[manuscript]{aastex}
\usepackage[utf8]{inputenc}
\usepackage{amsmath}
\usepackage{amssymb}
\usepackage{natbib}
\usepackage[flushleft]{threeparttable}
\usepackage{rotating}
\usepackage{rotating,booktabs}
\usepackage{longtable}

\graphicspath{{./}{figures/}}

\begin{document}

\title{Debris disks among \textit{Kepler} solar rotational analog stars}

\author{R. Silva Sobrinho\altaffilmark{1}, A. D. Da Costa\altaffilmark{1,2}, B. L. Canto Martins\altaffilmark{1,3}, I. C. Le\~ao\altaffilmark{1}, D. Freire da Silva\altaffilmark{1}, M. A. Teixeira\altaffilmark{1}, M. Gomes de Souza\altaffilmark{1}, D. B. de Freitas\altaffilmark{6}, J. P. Bravo\altaffilmark{1,4}, M. L. Das Chagas\altaffilmark{5}, and J. R. De Medeiros\altaffilmark{1}}

\affil{\altaffilmark{1}Departamento de F\'isica Te\'orica e Experimental, Universidade Federal do Rio Grande do Norte, Natal, RN 59072-970, Brazil.}
\affil{\altaffilmark{2}Universidade da Integra\c{c}\~ao da Lusofonia Afro-Brasileira, Reden\c{c}\~ao, CE 62790-000, Brazil.}
\affil{\altaffilmark{3}Observatoire de Gen\`eve, Universit\'e de Gen\`eve, Chemin des Maillettes, 51, Sauverny CH - 1290, Switzerland.}
\affil{\altaffilmark{4}Instituto Federal de Educa\c{c}\~ao, Ci\^encia e Tecnologia do Rio Grande do Norte, Natal, RN 59015-000, Brazil.}
\affil{\altaffilmark{5}Faculdade de F\'isica -- Instituto de Ci\^encias Exatas, Universidade Federal do Sul e Sudeste do Par\'a, Marab\'a, PA 68505-080, Brazil.}
\affil{\altaffilmark{6}Departamento de F\'isica, Universidade Federal do Cear\'a, Campus do Pici, Fortaleza, CE 60455-900, Brazil.}

\email{rodrigosobrinho@dfte.ufrn.br}

~\linebreak

\begin{abstract}
	
	Observations of circumstellar disks provide a powerful tool for our understanding of planetary systems dynamics. Analogs to the Solar System asteroid belts, debris disks result from the collision of the remaining solid material of the planet formation process. Even if the presence of disk is now reported for hundreds of stars, its detection around stars similar to the Sun is still very sparse. We report the results of a search for debris disks around \textit{Kepler} stars with surface physical parameters close to solar values, including rotation period, using observations by the Wide-field infrared Survey Explorer (\textit{WISE}). From the entire sample of {\it Kepler} stars, 881 targets were identified with these parameters and only six of them (KIC 1868785, 7267949, 7435796, 10533222, 11352643, and KIC 11666436) show unambiguous infrared excess, for which we determined debris disk physical parameters. Interestingly, the present study reveals traces of debris disks much more massive and brighter than the Solar System zodiacal dust, probably resulting from recent violent collisional events, orbiting stars with ages around the solar values. 

\end{abstract}


\section{INTRODUCTION}

The asteroid belt in our Solar System is located between Mars and Jupiter, i.e., between the inner terrestrial planets and the outer giant planets, with components presenting a large compositional diversity in size and distance from the Sun \citep{DeMeo}. It contains millions of irregularly shaped bodies composed of rocks, ices and metals with a total mass of approximately 4 percent of the Moon or 22 percent of that of Pluto. The presence of water vapor on Ceres, the largest body in the asteroid belt, and the identification of objects exhibiting apparently cometary activity yet orbiting completely within the main asteroid belt \citep{Hsieh 2006} are the most outstanding recent discoveries related to this region of the Solar System. Observations indicate that at the planetesimal formation stage the location of the snow line, which denotes the radius outside of which ice forms, was within the asteroid belt (e.g., \citealt{Martin2012}). Indeed, previous studies claimed that the inner asteroids, closest to Earth, at a radius of about 2.2 AU, were water devoid, whereas the outer asteroids, within a radius around 3.2 AU, were icy objects \citep{Abe 2000}. However, more recent studies indicate that asteroids are less segregated by water content than previously believed \citep{DeMeo}.

Although the presence of debris disks, with asteroid belt characteristics, is now well established for hundreds of stars (\citealt{chen06}; \citealt{cotten16}; \citealt{trilling08}; \citealt{Weissman 1995}; \citealt{aumann84}; \citealt{Patel14}), the present day literature pints for a scarcity of asteroid belt signatures around Sun-like stars (\citealt{Da Costa 2017}; \citealt{Sibthorpe 2018}). For instance, a recent study has shown a null detection of warm debris around solar twin stars \citep{Da Costa 2017}. Given this reality, we report here a search for infrared (IR) excess, a well established diagnostic for circumstellar debris disks, in a sample of 881 \textit{Kepler} main-sequence stars, using observations carried out with the \textit{Wide-field Infrared Survey Explorer} (WISE) \citep{Wright10}. This space mission mapped the sky at wavelengths 3.4, 4.6, 12, and 22 $\mu$m, known as filters W1, W2, W3, and W4, offering a unique laboratory for the search for stellar mid-IR excess. The 12 and 22 $\mu$m wavelengths are very sensitive to thermal emissions from objects at temperatures comparable to the Earth, around 300 K, and to the Solar System asteroid belt and interior zodiacal cloud, around 150-250 K.

\begin{table*}
	\centering
	\caption{Stellar parameters and fundamental disk properties for the Sun and the stars with confirmed IR excess. Stellar parameters: effective temperature ($T_{*}$), surface gravity (log $g$), metallicity ($[Fe/H]$), were obtained from \cite{huber14}, rotation periods ($P_{rot}$) were taken from \cite{McQuillan 2014}. Disk properties (this work): temperature of the dust ($T_d$), radius of the debris disk ($R_d$), dust mass of the disk of circumstellar material ($M_d$), the fractional luminosity of the dust ($f_d$).}
	\label{Table_1}
	\center
	\scriptsize
	
	\begin{tabular}{ccccccccc}
		\noalign{\smallskip}
		\hline 
		\hline
		
		\noalign{\smallskip}
		
		KIC & $T_{*}$ & {log ${g}$} & {[Fe/H]} & $P_{rot}$ &$T_d$ & $R_d$ & $M_d$ &  $f_d$  \\
		& (K) & (dex) & (dex) & (days)&(K) & (AU) & ($M_{\oplus}\times10^{-5}$) &  $(\times10^{-4})$  \\
		\hline
		\noalign{\smallskip}
		
		Sun 	&	5777$^{(a)}$	&	4.44$^{(b)}$	&	0.00	& 23.0--33.5$^{(c)}$		&	 276$^{(d)}$ 	&	 $< 5.0$$^{(d)}$ 	&	 {$\sim \times 10^{-4}$ $^{(d)}$} 	&	 { $\sim$ $10^{-3}$$^{(d)}$ } \\
		1868785	&	 5837$\pm$166 	&	 {4.50$\pm$0.29} 	&	 {-0.16$\pm$0.32} 	&	 24.219$\pm$0.232	&	 484 $\pm$22 	&	 0.33 $\pm$0.03 	&	 $2.23\pm 0.57$   	&	 $19.49\pm3.08$ \\
		7267949	&	 5629$\pm$159 	&	 {4.42$\pm$0.13} 	&	 {-0.42$\pm$0.36} 	&	 25.109$\pm$0.457 	&	 293 $\pm$13 	&	 0.77 $\pm$0.12 	&	 $5.36\pm1.98$ 	&	 $8.26 \pm1.77$ \\
		7435796	&	 5902$\pm$170 	&	 {4.43$\pm$0.06} 	&	 {0.14$\pm$0.18} 	&	 29.217$\pm$0.362	&	 280 $\pm$13 	&	 1.07 $\pm$0.12	&	 $12.16\pm 4.21$ 	&	 $9.60 \pm2.21$ \\
		10533222	&	 5926$\pm$176 	&	 {4.29$\pm$0.16} 	&	 {-0.12$\pm$0.26} 	&	 24.338$\pm$0.528	&	 341 $\pm$15 	&	 0.82 $\pm$0.15 	&	 $12.32 \pm4.66$ 	&	 $16.63\pm3.20$ \\
		11092105	&	 5658$\pm$162 	&	 {4.52$\pm$0.04} 	&	 {-0.20$\pm$0.28} 	&	 25.808$\pm$0.429	&	 265 $\pm$12 	&	 0.90 $\pm$0.09 	&	 $23.41\pm7.95$ 	&	 $26.41\pm6.23$ \\
		11666436	&	 5604$\pm$155 	&	 {4.56$\pm$0.03} 	&	 {-0.20$\pm$0.30} 	&	 23.923$\pm$0.579	&	 336 $\pm$15 	&	 0.52 $\pm$0.05 	&	 $4.07\pm1.18$ 	&	 $13.48\pm2.62$ \\
		\hline
		
	\end{tabular}
	\begin{minipage}{17cm}
		{References: {\em (a)} -  \citet{Neckel86}, {\em (b)} -  \citet{gray92}, {\em (c)} -  \citet{Lanza03}, {\em (d)} -  \citet{roberge2012}}
		
	\end{minipage}
	
\end{table*}

Indeed, thanks the high quality of the \textit{Kepler} photometric data, we are now able to study a new type of solar analog stars, the solar rotational analogs, namely those stars presenting atmospheric solar parameters and rotation periods similar to the Sun. The stellar sample here analyzed presents these unique characteristics, surface physical properties similar to the Sun and rotation period ranging $P_{rot}$ from 23 to 33 days. In section 2 of this Letter, we describe the WISE and \textit{Kepler} data used in this study. Section 3 describes the methods used in our analysis of these data. Finally, in Section 4, we present our results and discuss their implications.

\section{STELLAR WORKING SAMPLE AND WISE DATA ANALYSIS}

\subsection{The stellar sample}

For the present study, we use a sample of 881 \textit{Kepler} main-sequence stars with surface physical properties close to solar values, that is effective temperature in the range $5579\ K < Teff < 5979\ K$, superficial gravity in the range $3,94 \ cm s^{-2} < log\ g < 4,94\ cm s^{-2}$, metallicity [Fe/H] $\sim$ 0 and rotation period $P_{rot}$ from 23 to 33 days, namely the range of values of the Sun rotation period. Indeed, we have followed the same strategy by \citet{Das Chagas 2016}, with rotation period $P_{rot}$ taken from \citet{McQuillan 2014}.

The \textit{Kepler} coordinates of each target were then used to crosscheck with the 2MASS \citep{Cutri 2003} and full AllWISE \citep{cutri13} catalogs. Assuming a positional accuracy of 5 arcsecond, we find 862 stars with photometry in the three bands of 2MASS (J, H, and K) and in the four WISE bands (W1, W2, W3 and W4). The values of W3 and W4 magnitudes, SNRW3 and SNRW4 signal-to-noise ratios, and the confusion condition flag (ccf), were used as criteria to assess the quality and reliability of WISE data. Checking these photometric properties, we identified 447 stars with fundamental problems such as artifacts contamination (ccf=H,h,P,p,D,d and O,o) \citep{cutri13}, high saturation levels ($W3 < 3.8 $\ or\ $ W4 < -0.4$) and very low signal-to-noise ratio ($SNRW3/W4 < 2.0$). We have therefore disregarded these targets from our sample. Thus, a primary sample of 415 stars with non-saturated photometry, signal-to-noise greater than 2, and unaffected by known artifacts at one or both W3 and W4 bands was analyzed in the search for IR excess only in the band(s) in which no mentioned problem is found.

\subsection{Searching for IR excess}

Then, the observed spectral energy distribution (SEDs) and model-derived photospheric IR fluxes for each one of the referred 415 stars were compared using the Virtual Observatory Spectral Analyzer (VOSA, \citealt{bayo08}). The SEDs were constructed using the four IR bands W1-W4 from WISE \citep{cutri13}, the J, H, and Ks bands from 2MASS \citep{Cutri 2003}, and when available, the UBV bands \citep{Mermilliod06}, the G-band from Gaia \citep{van Leeuwen 2017}, and the color bands ugriz from SDSS \citep{Abazajian 2009}. For increasing the reliability of IR excess measurements, the theoretical fluxes were computed using three grids of theoretical stellar spectra: Kurucz-ATLAS9 \citep{Castelli97}, BT-DUSTY \citep{Allard12}, and BT-NextGen (AGSS2009) \citep{Allard12}. These models were used to determine the best-fitting line for the observed data by the $\chi^2$ minimization. Only the stars presenting IR excess from the above three models were chosen as IR excess candidates, amounting to 47 stars (see Table 2 in the online data). We also adopted the estimation of interstellar extinction provided in the \textit{Kepler} database.

For quantification of the observed IR excess, we used the excess significance parameter $\chi_{\lambda}$ (\citealt{beichman06}; \citealt{Moor2006}), defined as follows:

\begin{equation}
\chi_{\lambda}=\frac{F_{\lambda,obs}-F_{\lambda,phot}}{\sqrt{\sigma_{\lambda,obs}^2+\sigma_{\lambda,cal}^2}},
\end{equation} where $F_{\lambda,obs}$ is the observed flux density and $F_{\lambda,phot}$ is the expected photospheric flux density; $\sigma_{\lambda,obs}$ corresponds to the uncertainties of $F_{\lambda,obs}$; $\sigma_{\lambda,cal}$ refers to the calibration uncertainties of the WISE data of 4.5 \% and 5.7 \% in the W3 and W4 bands, respectively \citep{jarret11}. Here, we consider as presenting IR excess only those stars for which $\chi_{\lambda} \geq 2$ \citep{ribas12}, corresponding to at least 1.5$\sigma$ or 87 \% significance of deviation from photosphere IR emission ($\chi_{\lambda}= 0.0$). Based on this criterion, we find a total of 51 stars showing WISE mid-IR excess, although, only 47 stars present such excess in the three theoretical models, as explained before. 
The difference in significance between the Kurucz and the two other models fluctuates around 10\% for the W3 band and 1\% for the W4 band. This fluctuation gives us an order of magnitude of the systematic errors associated to the different physical ingredients considered in each model.
Our criterion of selecting objects that present excess simultaneously using different models should avoid a bias associated to a particular one \citep[e.g.,][]{sinclair2010}. Statistical errors due to the uncertainty in the fit with the models were not taken into consideration because they are negligible. \citet{Maldonado2017} point out that, for bright objects, the WISE calibration error is dominant in relation to the photometric and theoretical errors. This is not the case for our sample, composed of faint objects, where the photometric error is dominant over to the WISE calibration errors. The values of excess significance and the WISE coordinates for such stars are listed in Table 2 of the online data.

\subsection{WISE image inspection}

In order to identify which stars have a reliable IR excess, with no artificial artifacts or contamination, we applied to the sample of 47 stars the same procedure used by \citet{Da Costa 2017} for a visual inspection of the WISE images, based on the identification of some significant problems as PSFs (Point Spread Function) deformed due to an object close to the source, an absent or no evident object, or even caused by nearby objects blended, leading to a misinterpretation of the image. The WISE images were obtained from IRSA (Infrared Science Archive), using 0-3$\sigma$ linear scales around $1.7'\times1.7'$ of each IR excess candidate. In addition, we checked if the IR excess source is a punctual (circular) or elliptical (non-circular) or extensive object, we used a roundness criterion based on \citep{cotten16}, which consists on a comparison of bilateral symmetry of each source determined by a two-dimensional Gaussian adjustment defined by
\begin{equation}
Roundness\propto \frac{(\sigma_{x}-\sigma_{y})}{\frac{(\sigma_{x}+\sigma_{y})}{2}},
\end{equation} where $\sigma_x$ and $\sigma_y$ are the standard deviations of the Gaussian fit in the $x$ and $y$-axes, respectively. We tested several stars from the literature (e.g., \citealt{cotten16}) and, as a result, we concluded that those targets considered circular have the roundness smaller than 0.12, whereas non-circular objects have larger values. Applying this threshold to our sample, only 6 targets KIC 1868785, KIC 7267949, KIC 7435796, KIC 10533222, KIC 11092105 and KIC 11666436, survived all the image inspection criteria. The SEDs of the KIC 1868785, KIC 7267949, KIC 7435796, KIC 10533222, KIC 11092105 and KIC 11666436 stars, and their WISE images are shown in Figures 1 and 2, respectively.

\pagebreak
\section{Results}

Assuming that the detected IR excess in the stars KIC 1868785, KIC 7267949, KIC 7435796, KIC 10533222, KIC 11092105 and KIC 11666436 is related to IR radiation emitted by circumstellar dust, we modeled such excess using a simple blackbody function. A grid of blackbody temperatures ranging from 50 to 500 K at interval steps of 5 K (and 1 K, when necessary) was created, and performed the $\chi^2$ minimization to choose the dust temperature $T_d$ best fitting the observed excess, from where we obtained a temperature range of 265-484 K for the circumstellar dust.

In addition, we estimated three main dust properties, the fractional luminosity $f_d$, the dust radius $R_d$ and the dust mass $M_d$. The fractional dust luminosity, defined as the ratio of the luminosity from the dust to that of the star, was estimated using the relationship between dust and stellar fluxes given by \citet{beichman05}; considering that the detected excess for KIC 1868785, KIC 7267949, KIC 7435796, KIC 10533222, KIC 11092105 and KIC 11666436 has a peak emission at 12 $\mu$m wavelength, the minimum fractional luminosity was estimated, using values equals of 484 K, 293 K, 280 K, 341 K, 265 K and 336 K, for the dust temperature around each target, respectively, corresponding to the radiative temperature of a blackbody at 12 $\mu$m.

The disk radius or orbital $R_d$ was computed considering the dusty material as optically thin and in thermal equilibrium with the stellar radiation field. With these considerations, and assuming that the dust grains behave similarly to a blackbody, a minimum distance for the circumstellar dust was estimated following the recipe by \citet{backman93}. Finally, for the disk mass $M_d$ estimation we have applied the recipe given by \citet{liu14}. The computed debris disks physical parameters, temperature of the dust, radius of the debris disk, total mass of the disk of circumstellar material and the fractional luminosity of the dust are given in Table 1 together with stellar parameters.

\begin{figure}[h!]
	\includegraphics[scale=0.431]{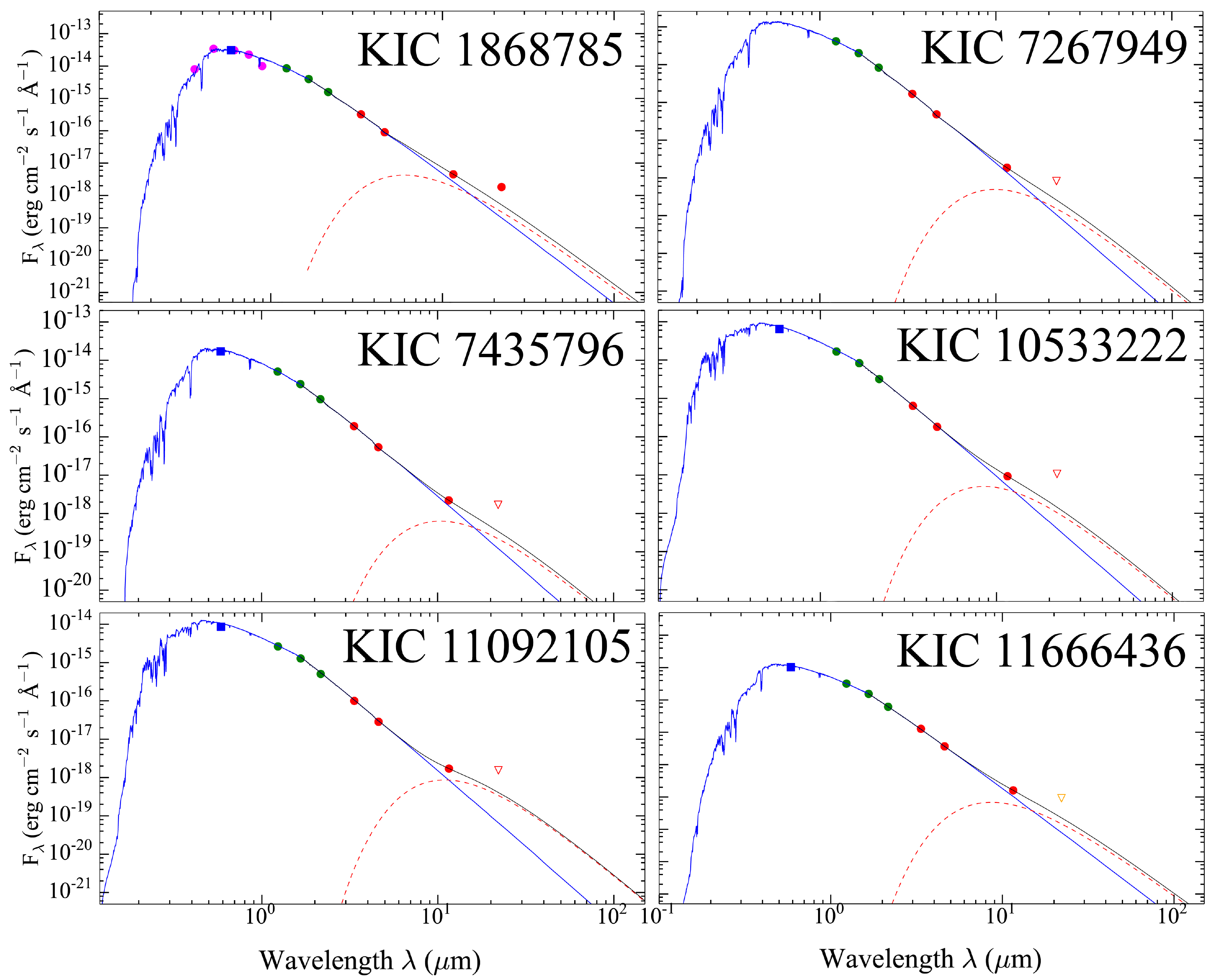}
	\caption{SEDs for stars with excess IR confirmed. The blue square represents the GAIA G-band flux \citep{van Leeuwen 2017}. Magenta, red and green circles indicate the photometric data points from SDSS (ugriz filter) \citep{Abazajian 2009}, 2MASS (JHK passbands) \citep{Cutri 2003}, and WISE \citep{cutri13}, respectively. The blue solid line denotes the adjustment of the KURUCZ model \citep{Castelli97}; the red dashed line means the best-fit using a simple blackbody model for WISE bands with IR excess and the black solid line is the sum of the two components.} 
	\label{Figure 1}
\end{figure}

\begin{figure}[h!]
	\centering
	\includegraphics[scale=1.3]{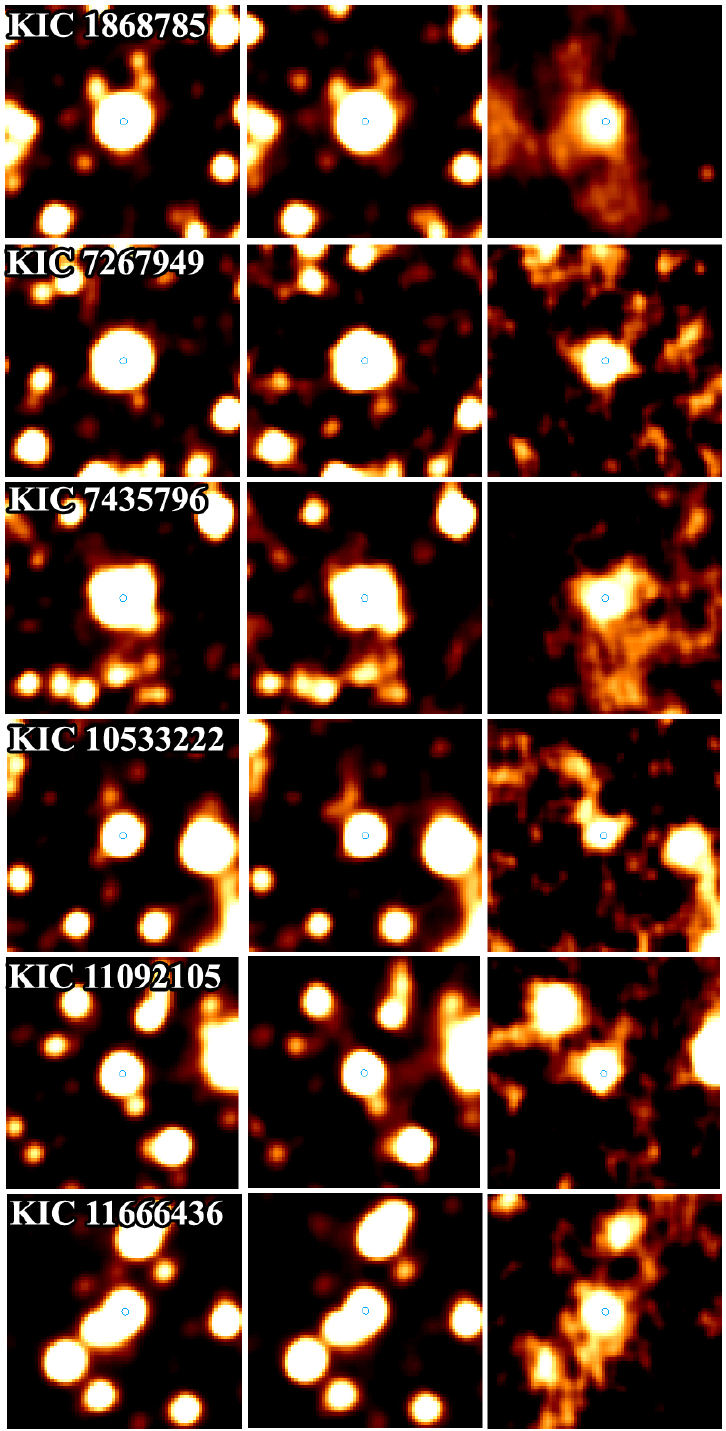}
	\caption{WISE images (from left to right: W1, W2, W3) for the six stars with IR excess confirmed in 12 $\mu$m, obtained from the Infrared Science Archive (IRSA) using 0-3$\sigma$ linear scales around $1.7'\times1.7'$. No contamination by artifacts has been found in the respective bands.} 
	\label{WISE_Imagens}
\end{figure}

\section{Conclusions}


This study reports the discovery of 6 \textit{Kepler} main-sequence stars, typically solar rotational analogs with rotation period similar to the Sun's values, presenting mid-IR excess compatible with the presence of debris disks. The dust temperatures obtained from the modeled IR excess range from 265 to 484 K, suggesting the presence of warm circumstellar material.
The computed dust parameters shows that the detected disks are located between 0.33 and 1.07 AU, at smaller orbital radii than the Solar System asteroid belt, that is from 2.0 and 3.5 AU
\citep{wyatt08}. 
Computed temperatures indicate that the referred stars with IR excess present warm circumstellar dust with temperatures, in average, higher than the solar asteroid belt value.
In effect, our finding may represent also an observational bias by considering that the presence of disks closer to the stars are hotter, and as a consequence, brighter if the IR excess is observed near the peak of their SEDs.

Circumstellar dust belts around main-sequence stars, as those reported in the present study, are composed of second-generation dust originated from the small-body population of planetary systems \citep{backman93}, which are mostly remnants of primordial protoplanetary disks \citep{Hernandez 2007}. These bodies can give fundamental informations about the chemistry and evolution of protoplanetary disk and the planetary systems they form. Despite a similar physical mechanism to be expected in the production of the reported debris disks, our findings show that stars with physical parameters similar to the Sun, as is the case of the whole sample here analyzed, can in fact be very different from the Sun once the star and its circumstellar environment are considered, confirming previous results by \cite{Da Costa 2017}. Among these physical parameters, age is an important one for determining the presence of debris disks. In the present work, based on gyrochronology estimations \citep[e.g.,][]{Barnes2016,Ceillier2016}, the stellar ages for our {\em Kepler} stars range around the solar age value, even though that range may be somewhat broad.

At the WISE wavelength bands, we are observing the Wien-edge of the energy distributions. In this sense, the lack of an excess for the large majority of the analyzed stars does not necessarily imply the absence of circumstellar material. Indeed, the detection of IR-excess, only in W3 band, is in agreement with \cite{liu14} assumption that the disk associated to this IR-excess are geometrically thin, that is, confined within a small radius range, with all the dust at the same temperature. The disks here reported may in fact be spatially extended and, by consequence, similar to the Solar System asteroid belt geometry. Their thin appearence may reflect the fact that only the inner edge of the disk can be detected with the present sensitivity. In addition, the abscence of detection in the W4 band may be explained by its considerably narrower range in comparison to W3. This fact would indicate that the fraction of solar rotational analog stars possessing debris disks could be higher than the fraction here observed. Nevertheless, the discovered debris disks are, by far, brighter and more massive than the Solar System zodiacal dust, a characteristic that allowed their detection. In this sense, the observation of solar debris disks at the distance of the refereed stars would be well below the WISE sensitivity level.

The present sample of debris disks has luminosity too high to be explained by a steady-state collisional cascade (\citealt{wyatt08}; \citealt{Gaspar2013}) and a large amount of warm dust that cannot be sustained at the estimated stellar ages \citep{wyatt08}. These unusual characteristics may reflect a possible disc-sculpting mechanism resulting from violent collisional events \citep[e.g.,][]{Kral2015,Kenyon2005, Kenyon2006,Raymond2009,Zappala2002,Durda2007}.  

Dust belts cooler than those reported here have their imprints at longer wavelength bands, and slight or no excess in the mid-IR. Therefore, the null detection of IR excess, at WISE sensitivity level, for the remaining 875 solar rotational analogue stars will certainly motivate new observational studies at far-IR, submillimeter and millimeter wavebands for a better characterization of material around these stars with a rotation period similar to that of the Sun. Furthermore, the presence of other disks structures \citep{wyatt08}, in particular cold components like the \textit{Kuiper} belt, and water ice traces, can be determined from observations in longer IR wavebands. In this sense, further observational studies are mandatory for the stars with detected IR excess here announced.

 \subsection{Acknowledgments}
 
 Research activity of the Observational Astronomy Board of the Federal University of Rio Grande do Norte (UFRN) is supported by continuous grants from CNPq and FAPERN Brazilian agencies. We also acknowledge financial support from INCT INEspa\c{c}o/CNPq/MCT. A.D.C. acknowledges a CAPES/PNPD fellowship. I.C.L acknowledges a CNPq/PDE fellowship. R.S.S., D.F.S., and M.N., acknowledge graduate fellowships from CAPES. This work is based on data products from the Wide-field Infrared Survey Explorer, a joint project of the University of California, Los Angeles, and the Jet Propulsion Laboratory/California Institute of Technology, supported by the National Aeronautics and Space Administration. This study has used NASA Astrophysics Data System (ADS) Abstract Service, the SIMBAD database, operated at CDS, Strasbourg, France, and data products from the Two Micron All-Sky Survey (2MASS), a joint project of the University of Massachusetts and the infrared Processing and Analysis Center, supported by the National Aeronautics and Space Administration and the National Science Foundation. This study has used VOSA support, developed under the Spanish Virtual Observatory project funded by the Spanish MICINN through grant AyA2011-24052. This study also includes data collected by the Kepler space mission. Funding for the Kepler mission is provided by the NASA Science Mission Directorate. We warmly thank the anonymous Reviewer for providing very helpful comments and suggestions.

\newpage


\section*{Online data}

\begin{table*}[h!]
	\centering
	\label{}
	\caption{List of the 51 stars with IR excess for which $\chi_{12}\geq 2.0$ or $\chi_{22}\geq 2.0$ from three photospheric models: Kurucz-ATLAS9, BT-DUSTY, and BT-NextGen. The ellipsis (\ldots) indicates upper limits for the WISE measurements.}
	\scriptsize
	
	\begin{tabular}{cccrccccccc}
		\noalign{\smallskip}
		\hline 
		\hline
		
		\noalign{\smallskip}
		
		KIC & RA & DE & \multicolumn{2}{c}{Kurukz} & \multicolumn{2}{c}{NextGen} & \multicolumn{2}{c}{BT-Dust} & \multicolumn{2}{c}{excess?} \\
		\hline
		\noalign{\smallskip}
		
		\ & \ & \ & $\chi_{12}$ &$\chi_{22}$& $\chi_{12}$ & $\chi_{22}$ & $\chi_{12}$ & $\chi_{22}$ & W3 & W4 \\
		1868785 & 291.57443 & 37.31554 & 6.08 & 3.20 & 5.75 & 3.18 & 5.75 & 3.18 & Yes & Yes \\
		3219127 & 286.57025 & 38.39762 & 2.16 &... & 2.01 &... & 2.01 &... & Yes & No \\
		3643036 & 290.73413 & 38.71746 & 3.22 & 2.92 & 3.11 & 2.91 & 3.11 & 2.91 & Yes & No \\
		4751492 & 293.19486 & 39.88782 & 2.20 &... & 2.11 &... & 2.11 &... & Yes & No \\
		4820062 & 286.56364 & 39.96158 & 2.20 &... &... &... &... &... & No & No \\
		4947417 & 297.54838 & 40.06008 & 6.21 &... & 6.18 &... & 6.18 &... & Yes & No \\
		5035733 & 297.72842 & 40.12920 & 9.60 & 8.77 & 9.53 & 8.77 & 9.53 & 8.77 & Yes & Yes \\
		5036092 & 297.79665 & 40.18130 & 4.21 &... & 4.03 &... & 4.03 &... & Yes & No \\
		5120654 & 297,01951 & 40,21359 & 9,93 & 4,87 & 9,71 & 4,86 & 9,71 & 4,86 & Yes & Yes \\
		5198141 & 294.87102 & 40.36343 & 2.27 &... & 2.12 &... & 2.12 &... & Yes & No \\
		5263998 & 288.34694 & 40.43668 &... & 2.05 &... & 2.06 &... & 2.06 & No & Yes \\
		5293988 & 296.67798 & 40.41841 & 6.09 &... & 6.01 &... & 6.01 &... & Yes & No \\
		5428470 & 284.34673 & 40.63319 & 2.37 &... & 2.19 &... & 2.19 &... & Yes & No \\
		5534914 & 292.24299 & 40.74677 &... & 2.41 &... & 2.41 &... & 2.41 & No & Yes \\
		5730371 & 298.05585 & 40.96706 & 11.19 & 7.04 & 11.10 & 7.03 & 11.10 & 7.03 & Yes & Yes \\
		5956717 & 290.62853 & 41.29013 & 2.09 &... & 2.15 &... & 2.15 &... & Yes & No\\
		6064473 & 297.64333 & 41.30418 & 4.62 & 5.83 & 4.62 & 5.83 & 4.62 & 5.83 & Yes & Yes \\
		6142317 & 296.86628 & 41.44202 &... & 2.12 &... & 2.12 &... & 2.12 & No & Yes \\
		6345900 & 284.64141 & 41.75273 & 2.06 &... &... &... &... &... & No & No \\
		6358701 & 290.02545 & 41.72927 & 2.49 &... & 2.43 &... & 2.43 &... & Yes & No \\
		6516101 & 289.85597 & 41.90501 & 4.83 & 2.01 & 4.70 & 2.01 & 4.70 & 2.01 & Yes & Yes \\
		6952979 & 293.17428 & 42.43163 & 3.50 &... & 2.83 &... & 2.83 &... & Yes & No \\
		

		\pagebreak[4]
		\global\pdfpageattr\expandafter{\the\pdfpageattr/Rotate 0}

	\end{tabular}
\end{table*}

\begin{table*}[h!]
	\centering
	\label{}
	\tablenum{2}
	\caption{\textit{Continued}}
	\center
	\scriptsize
	
	\begin{tabular}{cccrccccccc}
		\noalign{\smallskip}
		\hline 
		\hline
		
		\noalign{\smallskip}
		
		KIC & RA & DE & \multicolumn{2}{c}{Kurukz} & \multicolumn{2}{c}{NextGen} & \multicolumn{2}{c}{BT-Dust} & \multicolumn{2}{c}{excess?} \\
		\hline
		\noalign{\smallskip}
		
		\ & \ & \ & $\chi_{12}$ &$\chi_{22}$& $\chi_{12}$ & $\chi_{22}$ & $\chi_{12}$ & $\chi_{22}$ & W3 & W4 \\
		7267949 & 286.94895 & 42.82167 & 2.35 &... & 2.24 &... & 2.24 &... & Yes & No \\
		7435796 & 288.98419 & 43.03943 & 2.85 &... & 2.75 &... & 2.75 &... & Yes & No \\
		7457546 & 295.36482 & 43.03973 &... & 2.09 &... & 2.08 &... & 2.08 & No & Yes \\
		7763388 & 295.18879 & 43.42369 & 2.20 &... & 2.06 &... & 2.06 &... & Yes & No \\
		7772296 & 297.59851 & 43.44764 & 2.36 &... & 2.30 &... & 2.30 &... & Yes & No \\
		7953983 & 289.71354 & 43.72041 & 2.34 &... & 2.30 &... & 2.30 &... & Yes & No \\
		8042782 & 296.39621 & 43.85904 & 2.39 &... & 2.36 &... & 2.36 &... & Yes & No \\
		8328443 & 300.40546 & 44.26359 & 3.76 & 5.45 & 3.74 & 5.45 & 3.74 & 5.45 & Yes & Yes \\
		8429890 & 291.47401 & 44.48705 & 2.28 &... & 2.24 &... & 2.24 &... & Yes & No \\
		8441073 & 295.51828 & 44.44331 & 2.71 &... & 2.70 &... & 2.70 &... & Yes & No \\
		8565235 & 293.87219 & 44.64762 &... & 2.11 &... & 2.10 &... & 2.10 & No & Yes \\
		8718439 & 300.81718 & 44.82276 &... & 2.16 &... & 2.16 &... & 2.16 & No & Yes \\
		9161405 & 294.41346 & 45.55480 & 2.83 &... & 2.82 &... & 2.82 &... & Yes & No \\
		9410702 & 294.43354 & 45.91136 & 2.53 &... & 2.47 &... & 2.47 &... & Yes & No \\
		9896250 & 296.06648 & 46.72492 &... & 2.11 &... & 2.10 &... & 2.10 & No & Yes \\
		9946870 & 289.92768 & 46.85935 & 3.86 &... & 3.73 &... & 3.73 &... & Yes & No \\
		9963105 & 296.50230 & 46.87746 & 2.02 &... &... &... &... &... & No & No \\
		10468501 & 291.10030 & 47.66703 & 2.03 &... &... &... &... &... & No & No \\
		10533222 & 291.07269 & 47.71635 & 2.29 &... & 2.24 &... & 2.24 &... & Yes & No \\
		10613866 & 296.88037 & 47.81899 & 2.12 &... & 2.08 &... & 2.08 &... & Yes & No \\
		10670950 & 293.99346 & 47.95911 & 3.63 &... & 3.61 &... & 3.61 &... & Yes & No \\

		\pagebreak[4]
		\global\pdfpageattr\expandafter{\the\pdfpageattr/Rotate 0}

	\end{tabular}
\end{table*}

\begin{table*}[h!]
	\centering
	\label{}
	\tablenum{2}
	\caption{\textit{Continued}}
	\center
	\scriptsize
	
	\begin{tabular}{cccrccccccc}
		\noalign{\smallskip}
		\hline 
		\hline
		
		\noalign{\smallskip}
		
		KIC & RA & DE & \multicolumn{2}{c}{Kurukz} & \multicolumn{2}{c}{NextGen} & \multicolumn{2}{c}{BT-Dust} & \multicolumn{2}{c}{excess?} \\
		\hline
		\noalign{\smallskip}
		
		\ & \ & \ & $\chi_{12}$ &$\chi_{22}$& $\chi_{12}$ & $\chi_{22}$ & $\chi_{12}$ & $\chi_{22}$ & W3 & W4 \\
		
		10972628 & 290.27466 & 48.48056 & 3.51 &... & 2.86 &... & 2.86 &... & Yes & No \\
		11074641 & 286.59790 & 48.63996 &... & 2.13 &... & 2.13 &... & 2.13 & No & Yes \\
		11092105 & 295.56162 & 48.64493 & 4.05 &... & 4.02 &... & 4.02 &... & Yes & No \\
		11135275 & 290.67574 & 48.72052 &... & 2.19 &... & 2.19 &... & 2.19 & No & Yes \\
		11177716 & 284.26121 & 48.88640 & 2.47 &... & 2.37 &... & 2.37 &... & Yes & No \\
		11352643 & 293.02650 & 49.18330 & 3.40 &... & 3.31 &... & 3.31 &... & Yes & No \\
		11666436 & 294.26457 & 49.73296 & 2,64 &... & 2.29 &... & 2.29 &... & Yes & No \\
		12256697 & 290.46572 & 50.93251 & 2.98 & 2.20 & 2.80 & 2.19 & 2.80 & 2.19 & Yes & Yes \\
		\hline

		\pagebreak[4]
		\global\pdfpageattr\expandafter{\the\pdfpageattr/Rotate 0}

	\end{tabular}
\end{table*}

\newpage

\begin{table*}[h!]
	\centering
	\label{}
	\tablenum{3}
	\caption{Visual inspection of WISE images for a sample of 47 stars with IR exces.}\label{tabela_visual}
	\center
	\scriptsize
	
	\begin{tabular}{lcc}
		\noalign{\smallskip}
		\hline 
		\hline
		
		\noalign{\smallskip}
		
		KIC & visual inspection & Disk \\
		\hline
		\noalign{\smallskip}
		
		1868785 & Isolated single source in W3 & Yes \\
		3219127 & Non-circular sources in W3 & No \\
		3643036 & Offset source in W3 and W4 & No \\
		4751492 & Absent source in W3 & No \\
		4947417 & Absent source in W3 & No \\
		5035733 & Absent source in W3 and W4 & No \\
		5036092 & Offset source in W3 & No \\
		5120654 & Absent source in W3 and W4 & No \\
		5198141 & Confusion of sources in W3 & No \\
		5263998 & Offset source in W4 & No \\
		5293988 & Offset source in W3 & No \\
		5428470 & Non-circular source in W3 & No \\
		5534914 & Offset source in W4 & No \\
		5730371 & Absent source in W3 and W4 & No \\
		5956717 & Offset source in W3 & No \\
		6064473 & Absent source in W3 and W4 & No \\
		6142317 & Offset source in W4 & No \\
		6358701 & Offset source in W3 & No \\
		6516101 & Non-circular source in W3 and absent source in W4 & No \\
		6952979 & Confusion of sources in W3 & No \\
		7267949 & Isolated single source in W3 & Yes \\
		7435796 & Isolated single source in W3 & Yes \\
		7457546 & Absent source in W4 & No \\
		7763388 & Non-circular source in W3 & No \\

		\pagebreak[4]
		\global\pdfpageattr\expandafter{\the\pdfpageattr/Rotate 0}

	\end{tabular}
\end{table*}

\newpage

\begin{table*}[h!]
	\centering
	\label{}
	\tablenum{3}
	\caption{\textit{Continued}}
	\center
	\scriptsize
	
	\begin{tabular}{lcc}
		\noalign{\smallskip}
		\hline 
		\hline
		
		\noalign{\smallskip}
		
		KIC & visual inspection & Disk \\
		\hline
		\noalign{\smallskip}
		
		7772296 & Absent source in W3 & No \\
		7953983 & Offset source in W3 & No \\
		8042782 & Offset source in W3 & No \\
		8328443 & Absent source in W3 and W4 & No \\
		8429890 & Absent source in W3 & No \\
		8441073 & Absent source in W3 & No \\
		8565235 & Absent source in W4 & No \\
		8718439 & Absent source in W4 & No \\
		9161405 & Absent source in W3 & No \\
		9410702 & Non-circular source in W3 & No \\
		9896250 & Absent source in W4 & No \\
		9946870 & Non-circular source in W3 & No \\
		10533222 & Isolated single source in W3 & Yes \\
		10613866 & Offset source in W3 & No \\
		10670950 & Confusion of sources in W3 & No \\
		10972628 & Offset source in W3 & No \\
		11074641 & Absent source in W4 & No \\
		11092105 & Isolated single source in W3 & Yes \\
		11135275 & Offset source in W4 & No \\
		11177716 & Absent source in W3 & No \\
		11352643 & Non-circular source in W3 & No \\
		11666436 & Isolated single source in W3 & Yes \\
		12256697 & Non-circular source in W3 and absent in W4 & No \\
		\hline

		\pagebreak[4]
		\global\pdfpageattr\expandafter{\the\pdfpageattr/Rotate 0}

	\end{tabular}
\end{table*}

\newpage

\begin{table*}[h!]
	\centering
	\tablenum{4}
	\caption{Stellar parameters for the stars with confirmed IR excess. Effective temperature ($T_{*}$), surface gravity (log $g$), metallicity ($[Fe/H]$), radius ($R_*$) and mass ($M_*$) were obtained from \cite{huber14}, whereas luminosity ($L_*$) was computed using these parameters. Rotation periods ($P_{rot}$) were taken from \cite{McQuillan 2014}.}
	\label{quality_protometry}
	\center
	\scriptsize

	\begin{tabular}{cccccccc}
		\noalign{\smallskip}
		\hline 
		\hline
		
		\noalign{\smallskip}
		
		KIC & $T_{*} (K)$ & log ${g}$ & [Fe/H] & $R_{*}$($R_{\odot})$ & $M_{*}$ ($M_{\odot}$) & $L_{*}$ ($L_{\odot}$) & $P_{rot}$ (days) \\
		\hline
		\noalign{\smallskip}

		\hline
		1868785 & 5837$\pm$166 & 4.50$\pm$0.29 & -0.16$\pm$0.32 & 0.974$\pm$0.079 & 1.083$\pm$0.141 & 0.988$\pm$0.275 & 24.219$\pm$0.232 \\
		7267949 & 5629$\pm$159 & 4.42$\pm$0.13 & -0.42$\pm$0.36 & 0.900$\pm$0.131 & 0.775$\pm$0.049 & 0.729$\pm$0.373 & 25.109$\pm$0.457 \\
		7435796 & 5902$\pm$170 & 4.43$\pm$0.06 & 0.14$\pm$0.18 & 1.045$\pm$0.107 & 1.072$\pm$0.129 & 1.188$\pm$0.342 & 29.217$\pm$0.362 \\
		10533222 & 5926$\pm$176 & 4.29$\pm$0.16 & -0.12$\pm$0.26 & 1.176$\pm$0.209 & 0.976$\pm$0.107 & 1.530$\pm$0.537 & 24.338$\pm$0.528 \\
		11092105 & 5658$\pm$162 & 4.52$\pm$0.04 & -0.20$\pm$0.28 & 0.852$\pm$0.076 & 0.885$\pm$0.079 & 0.667$\pm$0.255 & 25.808$\pm$0.429 \\
		11666436 & 5604$\pm$155 & 4.56$\pm$0.03 & -0.20$\pm$0.30 & 0.815$\pm$0.064 & 0.887$\pm$0.088 & 0.587$\pm$0.222 & 23.923$\pm$0.579 \\
		\hline
		
	\end{tabular}

\end{table*}

\newpage


\figsetstart
\figsetnum{3}
\figsettitle{SEDs and WISE images of the excluded stars from our sample.}

\figsetgrpstart
\figsetgrpnum{3.1}
\figsetgrptitle{SEDs and WISE images of KIC 3219127, KIC 3643036, KIC 4751492, KIC 4947417, and KIC 5035733.}
\figsetplot{figure_3_online_data}
\figsetgrpnote{}
\figsetgrpend

\figsetgrpstart
\figsetgrpnum{3.2}
\figsetgrptitle{SEDs and WISE images of KIC 5036092, KIC 5120654, KIC 5198141, KIC 5263998, and KIC 5293988.}
\figsetplot{figure_4_online_data}
\figsetgrpnote{}
\figsetgrpend

\figsetgrpstart
\figsetgrpnum{3.3}
\figsetgrptitle{SEDs and WISE images of KIC 5428470, KIC 5534914, KIC 5730371, KIC 5956717, and KIC 6064473.}
\figsetplot{figure_5_online_data}
\figsetgrpnote{}
\figsetgrpend

\figsetgrpstart
\figsetgrpnum{3.4}
\figsetgrptitle{SEDs and WISE images of KIC 6142317, KIC 6358701, KIC 6516101, KIC 6952979, and KIC 7457546.}
\figsetplot{figure_6_online_data}
\figsetgrpnote{}
\figsetgrpend

\figsetgrpstart
\figsetgrpnum{3.5}
\figsetgrptitle{SEDs and WISE images of KIC 7763388, KIC 7772296, KIC 7953983, KIC 8042782, and KIC 8328443.}
\figsetplot{figure_7_online_data}
\figsetgrpnote{}
\figsetgrpend

\figsetgrpstart
\figsetgrpnum{3.6}
\figsetgrptitle{SEDs and WISE images of KIC 8429890, KIC 8441073, KIC 8565235, and KIC 8718439.}
\figsetplot{figure_8_online_data}
\figsetgrpnote{}
\figsetgrpend

\figsetgrpstart
\figsetgrpnum{3.7}
\figsetgrptitle{SEDs and WISE images of KIC 9161405, KIC 9410702, KIC 9896250, KIC 9946870, and KIC 10613866.}
\figsetplot{figure_9_online_data}
\figsetgrpnote{}
\figsetgrpend

\figsetgrpstart
\figsetgrpnum{3.8}
\figsetgrptitle{SEDs and WISE images of KIC 10670950, KIC 10972628, KIC 11074641, KIC 11135275, and KIC 11177716.}
\figsetplot{figure_10_online_data}
\figsetgrpnote{}
\figsetgrpend

\figsetgrpstart
\figsetgrpnum{3.9}
\figsetgrptitle{SEDs and WISE images of KIC 11352643 and KIC 12256697.}
\figsetplot{figure_11_online_data}
\figsetgrpnote{}
\figsetgrpend

\figsetend

\begin{figure}
	\figurenum{3}
	\epsscale{0.85}
	\plotone{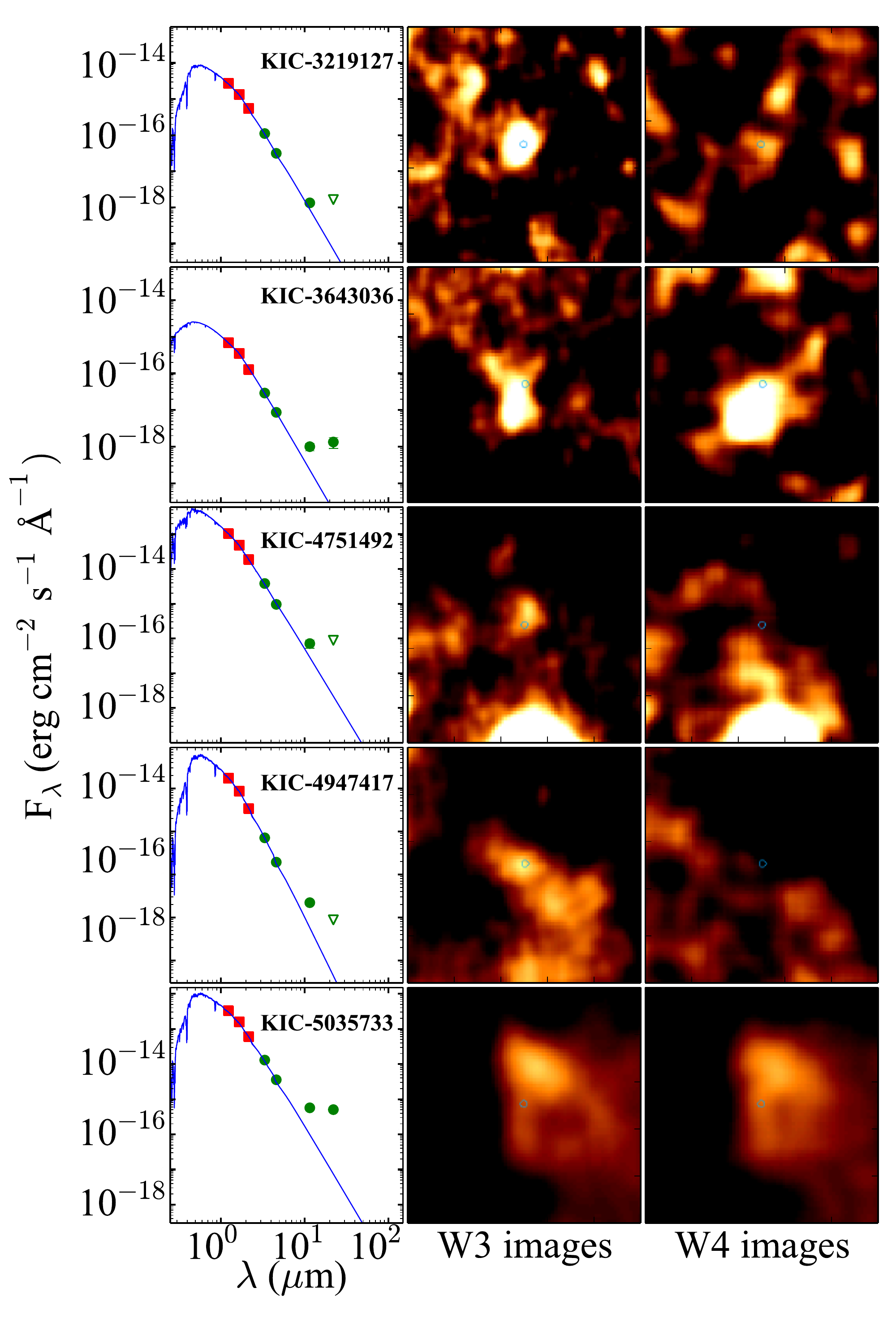}
	\caption{SEDs and WISE images of KIC 3219127, KIC 3643036, KIC 4751492, KIC 4947417, and KIC 5035733 stars with apparent IR excesses, but presenting fundamental problems in the images. \textit{Left panels}: SED of each individual target. Red squares represent the fluxes from the 2MASS JHK passbands \citep{Cutri 2003}. Green circles display the fluxes from WISE W1-W4 bands \citep{cutri13}. The WISE upper limits are indicated by green open triangles. The blue solid line denotes the adjustment of the Kurucz model \citep{Castelli97}. \textit{Center/Right panels}: respective stellar WISE W3/W4 images. The complete figure set (with 9 subsets comprising 41 targets) is available in the online journal.}
\end{figure}


\clearpage

\end{document}